Running head: INFORMANTS AND HERITABILITY

Informant Discrepancies and the Heritability of Antisocial Behavior: A Meta-Analysis

Elizabeth Talbott, George Karabatsos, & Jaime Zurheide

University of Illinois at Chicago

Address correspondence to:
Elizabeth Talbott, Ph.D.
Associate Professor
Special Education
University of Illinois at Chicago
1040 W. Harrison Street (MC 147)
Chicago, IL 60607-7133
Phone: 312 413-8745
Fax: 312 996-5651
Email: Etalbott@uic.edu

This research is supported by NSF Grant SES-1156372, from the program in Methodology, Measurement, and Statistics.

Running head: INFORMANTS AND HERITABILITY

Informant Discrepancies and the Heritability of Antisocial Behavior: A Meta-Analysis

ABSTRACT

Antisocial behavior, which includes both aggressive and delinquent activities, is the opposite of prosocial behavior. Researchers have studied the heritability of antisocial behavior among twin and non-twin sibling pairs from behavioral ratings made by parents, teachers, observers, and youth. Through a meta-analysis, we examined longitudinal and cross sectional research in the behavioral genetics of antisocial behavior, consisting of 42 studies, of which 38 were studies of twin pairs, 3 were studies of twins and non-twin siblings, and 1 was a study of adoptees. These studies provided $n = 89$ heritability ($h^2$) effect size estimates from a total of 94,517 sibling pairs who ranged in age from 1.5 to 18 years; studies provided data for 29 moderators (predictors). We employed a random-effects meta-analysis model to achieve three goals: (a) perform statistical inference of the overall heritability distribution in the underlying population of studies, (b) identify significant study level moderators (predictors) of heritability, and (c) examine how the heritability distribution varied as a function of age and type of informant, particularly in longitudinal research. The meta-analysis indicated a bimodal overall heritability distribution, indicating two clusters of moderate and high heritability values, respectively; identified four moderators that predicted significant changes in mean heritability; and indicated differential patterns of median $h^2$ and variance (interquartile ranges) across informants and ages. We argue for a cross-perspective, cross-setting model for selecting informants in behavioral genetic research, that is flexible and sensitive to changes in antisocial behavior over time.

Informant Discrepancies and the Heritability of Antisocial Behavior: A Meta-Analysis

Antisocial Behavior has been defined as a "repeated violation of social norms across a range of contexts," including home, school, and community (Walker, Ramsey, & Gresham, 2004, p. 3). It is the opposite of prosocial behavior and has been operationalized in three major categories: psychiatric diagnoses (Conduct Disorder, CD and Antisocial Personality Disorder, ASPD, Loeber, Burke, & Pardini, 2009); aggression and externalizing behavior (Achenbach & Edelbrock, 1978); and delinquency or rule-breaking behavior (Reid, Patterson, & Snyder, 2002; Rhee & Waldman, 2002).

Antisocial behavior shows strong continuity across development, with aggressive behavior emerging in early childhood and remaining stable throughout adulthood for approximately 5-10% of the population (Burt, 2009a; Dodge & McCourt, 2010; Moffitt, 1993; Stanger, Achenbach, & Verhulst, 1997; Tremblay, 2000). During late childhood and adolescence, patterns of rule-breaking and delinquent antisocial behavior begin to emerge and increase in frequency, following divergent paths and peaking during adolescence, both for those who participate in antisocial acts throughout life, and those who experiment as teens (Loeber et al., 1993; Moffitt, 1993). Although overall levels of aggression in the population decline from early childhood to adulthood, frequency of rule-breaking and delinquent behavior shows a steep increase over the course of adolescence (Burt, 2009a; Moffitt, 1993). Hence, the longitudinal study of antisocial behavior is vital.

**Antisocial Behavior and Informant Discrepancies**

Because antisocial behavior is likely to occur across settings (home, school, neighborhood, community), and in interactions with disparate individuals (parents, siblings, teachers, peers), diverse and multiple informants have been employed to assess child and adolescent antisocial behavior (Achenbach, McConaughy, & Howell, 1987; Patterson et al., 1992). In fact, obtaining the views of multiple informants about child and adolescent behavior is considered to be an evidence-based practice in the assessment of child and adolescent mental health (Hunsley & Mash, 2007; Mash & Hunsley, 2005).

    Yet is unclear the extent to which adding data from multiple informants improves prediction (Kraemer, Measelle, Ablow, Essex, Boyce, & Kupfer, 2003). This is because there is no single measure or method of assessing antisocial behavior in children that provides a definitive or gold standard (De Los Reyes & Kazdin, 2005; Kraemer et al., 2003). In fact, informants consistently disagree in their ratings of child and adolescent behavior, a finding that is robust and has persisted across assessment methods (i.e., rating scales and interviews), among individuals from diverse ethnic and cultural backgrounds, and in clinic and community samples (Achenbach et al., 1987; De Los Reyes & Kazdin, 2005). Different informants witness behavior in different settings and provide attributions for behavior from differing perspectives. This is not measurement error; instead, disagreements among informants likely reflect variation in informants' perspectives on behavior and the contexts for behavior (Achenbach, 2011; Kraemer et al., 2003). For example, parents and teachers might attribute a child's behavior to internal characteristics of the child, to his or her biology or disposition, children and adolescents might attribute their behavior to external characteristics associated with the environment and social contexts (De Los Reyes & Kazdin, 2005). In addition, specific informants may be only privy to behavior in particular settings, such as teenagers observing fellow teens engage in rule-breaking and teachers' observations of children and youth primarily in the classroom.

    One informant is not necessarily superior to the other; rather, informant discrepancies might be revealing. For example, discrepancies between parent and teacher ratings of disruptive

behavior can be linked to variations in laboratory observations of children's behavior (De Los Reyes, Henry, Tolan, & Wakschlag, 2009; Wakschlag et al., 2008). In addition, discrepant scores between parents and adolescents on the Achenbach measures (Achenbach & Rescorla, 2001) can predict a range of poor outcomes, including alcohol and substance use, contact with the police or judicial system, expulsion from school, firing from a job, unwanted pregnancy, attempts at self-harm, self-reports of having a behavioral or emotional problem, and referral to mental health services (Ferdinand, van der Ende, & Verhulst, 2004). Increasing the range of informants and aggregating their information has been the response to this unique measurement challenge, but it may not be the solution (Hartley, Zakriski, & Wright, 2011).

## Antisocial Behavior and Behavioral Genetics

Antisocial behavior is complex, follows diverse pathways, and changes across development (Loeber et al., 1993; Moffitt, 1993; Patterson, Reid, & Snyder, 2002). Like all complex social behavior, antisocial behavior is influenced by genes (Turkheimer, 1998; 2000). In fact, genetics play a major role in the risk that individuals have for a broad range of mental disorders, including antisocial behavior (Kendler, 2005a; 2005b; Malouff, Rooke, & Schutte, 2008; Rhee & Waldman, 2002; Rutter, Moffitt, & Caspi, 2006). The expression of genes on antisocial behavior over the lifespan is dynamic, malleable, and responsive to the social environment throughout development (Reiss & Neiderhiser, 2000).

Yet, even as research in behavioral genetics moves forward, from the work of partitioning variance into genetic and environmental components to identifying specific genes that may affect behavior (Turkheimer, 1998), findings still depend upon the ratings of individual informants (parents, teachers, observers, and youth) and data collected from official records (i.e., school and criminal records). Such ratings, describing child and adolescent behavior, carry with them all of the challenges associated with employing different informants, including the confounding of informant and age (Achenbach, 2011; Achenbach, et al., 1987; De Los Reyes, 2011; De Los Reyes & Kazdin, 2005; Rhee & Waldman, 2002). In addition, behavioral genetic research does not consistently rely on multiple informants who have diverse perspectives about behavior in different contexts.

In order to study the behavioral genetics of antisocial behavior, researchers have conducted studies at three levels: first, they have employed methods associated with basic genetic epidemiology (Kendler, 2005). That is, they collect informants' ratings of the behavior of pairs of siblings having a range of genetic relatedness, including twins (monozygotic and dyzogotic) as well as biological full, half, and unrelated siblings, with the purpose of ascertaining the extent to which the behavior or set of behaviors is heritable (Kendler, 2005; Reiss, Neiderhiser, Hetherington, & Plomin, 2000). Using the differences between siblings on measures of antisocial behavior, researchers partition the variance in behavior into genetic (both additive, $a^2$ and nonadditive, $d^2$) and environmental (shared, $c^2$ and nonshared plus error, $e^2$) components (Burt, 2009b). Additive and nonadditive genetic influences together comprise heritability, which is measured as the "ratio of genetic variance to total trait variance" (Johnson, Penke, & Spinath, 2011, p. 257). At the second level, using methods associated with advanced genetic epidemiology, researchers explore the nature and mode of action of genetic risk factors (Kendler, 2005).

At the third level, behavioral genetic researchers focus on gene finding, including the interaction between genetic loci and life experiences, or G x E interactions, and their effects on aggressive and delinquent behavior in adolescents and adults (Bernet, Vnencak-Jones, Farahany, & Montgomery, 2007). Even as researchers seek to move behavioral genetic research to the second and third levels, the third level being the study of specific genes interacting with environmental events to predict antisocial behavior (Moffitt, 2005), outcomes continue to be measured by different informants (i.e., Caspi et al., 2002; Kim-Cohen et al., 2006). Although a small number of individual studies have explored the convergence and divergence of different informants in the heritability of antisocial behavior (i.e., Baker, Jacobson, Raine, Lozano, & Bezdjian, 2007; Burt, McGue, Krueger, & Iacono, 2005; Simonoff et al., 1995), no single study has combined the results from behavioral genetic research to examine the contributions of different informants across ages. Such a study is needed to identify the fundamental questions about heritability as it has been derived from the ratings of parents, teachers, youth, and observers (Achenbach, 2011; De Los Reyes, 2011; De Los Reyes & Kazdin, 2005; Hartley, Zakriski, & Wright, 2011).

In the following sections, we review previous meta-analyses of antisocial behavior that have included age and informant, along with other study characteristics, as significant moderators of the heritability of antisocial behavior. Moderators examined in these studies have included participant characteristics (i.e., age, gender, and method of determining twin zygosity); antisocial construct measured (i.e., aggression, delinquency, externalizing behavior, or criminality); method of assessment (i.e., diagnostic interview, questionnaire); and type of informant (i.e., mother, father, teacher, observer, peer).

**Meta-analyses of Behavioral Genetic Studies of Antisocial Behavior in Children and Adolescents**

We identified five meta-analyses of heritability of antisocial behavior that included results by informant and age, both in cross-sectional and longitudinal research. For each of these meta-analyses, we report (a) overall heritability of antisocial behavior; (b) significant moderators of heritability; and (c) patterns of heritability by informant and age. In an early meta-analysis, Miles and Carey (1997) examined 20 twin studies and 4 adoption studies and found that approximately 50% of the variance in antisocial behavior was attributable to heritability. Miles and Carey (1997) examined several potential moderators of heritability, including informant (parent, self, observer), age, sex, sibling type (biological or adoptive), and zygosity (for twins). Miles and Carey (1997) found that heritability for males was higher than that for females, and that studies with parents as informants yielded lower heritability than studies with youth as informants, even as informant and age were confounded (Miles & Carey, 1997). That is, Miles and Carey (1997) were among the first to report that parents and observers were more likely to rate children, and adolescents and adults were more likely to rate themselves (Miles & Carey, 1997). These early findings were heuristic in their contributions to the future study and meta-analyses of antisocial behavior.

Four more recent meta-analyses explored heritability estimates for antisocial behavior among children and adolescents, significant moderators of heritability, and patterns of heritability across ages (Bergen, Gardner, & Kendler, 2007; Burt, 2009a; Burt, 2009b; Rhee & Waldman, 2002). Together, these meta-analyses have significantly advanced the field and contributed to the framing of the present study. First, our measure of the antisocial construct and selection of moderators mirrors those selected by Rhee & Waldman (2002) and Burt (2009a; 2009b). Second, our interest in the measure of heritability across ages reflects the approach taken by Bergen et al. (2007) and builds on that research by adding a larger pool of studies. And third, the finding by Rhee & Waldman (originally identified by Miles and Carey, 1997) that age and informant were confounded, followed by the reporting of heritability separately for age and informant by Burt (2009b), prompted our effort to unpack age and informant in the assessment of heritability, with a focus on data from longitudinal samples, and controlling for all moderators in the model.

Rhee and Waldman (2002) conducted a comprehensive meta-analysis of the magnitude of genetic and environmental influences on antisocial behavior using correlations from 42 independent twin and 10 independent adoption samples from 51 studies, and examining the possible moderating influences of individual characteristics (i.e., age and gender of participants); study characteristics (i.e., operationalization of the antisocial construct and method of assessing antisocial behavior, as well as zygosity determination method) on genetic and environmental influences. Rhee and Waldman (2002) reported moderate proportions of variance in antisocial behavior attributed to genetic influences ($a^2 = .41$), and found that age, zygosity determination method, operationalization of the antisocial construct, and informant of antisocial behavior were all significant moderators of the magnitude of genetic and environmental effects (Rhee & Waldman, 2002). In addition, Rhee and Waldman (2002) found that studies using reports by others (parent, teacher, and observer ratings of children) yielded higher estimates of heritability ($a^2 = .53$) than studies using reports by adolescents themselves ($a^2 = .39$), and by studies using the results of criminal records ($a^2 = .33$). Furthermore, type of informant was confounded with age: parents and teachers consistently reported on behavior in childhood, whereas adolescents reported on their own behavior (Rhee & Waldman, 2002).

Burt (2009b) conducted a meta-analysis of shared environment influences ($c^2$) in externalizing behavior ($n = 16$ studies) and conduct problems ($n = 38$ studies) in children and adolescents, with separate analyses by sex, informant, and age. Burt (2009b) also provided the results for heritability by informant and age, which we review below. Burt (2009b) employed definitions of externalizing and conduct problems that match our definitions of the constructs externalizing/antisocial and CD, and reported moderate-to-high heritability for externalizing and conduct problems (averaged), $a^2 = .583$.

Furthermore, Burt (2009b) reported heritability for externalizing and conduct problems (averaged here) among children and adolescents separately by informant and age. Across the two antisocial constructs, heritability for ratings completed by others (mothers, teachers, and fathers, averaged) was $a^2 = .561$, where heritability for ratings completed by youth was $a^2 = .431$. These were comparable to those findings reported by Rhee & Waldman (2002). Similarly, Burt (2009b) reported heritability for externalizing and conduct problems (averaged here), depending upon the age at which the construct was assessed. Burt (2009b) identified heritability of $a^2 = .569$ at ages 1-5; $a^2 = .602$ at ages 6-10; and $a^2 = .538$ at ages 11-18.

In further exploration of the genetic influences on antisocial behavior, Burt (2009a) examined whether aggressive and rule-breaking (delinquent) forms of antisocial behavior differed in patterns of genetic and environmental influence. In a meta-analysis of 19 studies of aggressive behavior and 15 studies of non-aggressive behavior (with sex, informant, and age as moderators), Burt (2009a) found aggressive behavior ($a^2 = .651$) to be more highly heritable than rule-breaking ($a^2 = .481$). In subsequent analyses, Burt (2009a) sought to determine whether these differences in heritability persisted across informant and age.

Burt (2009a) found that heritability for aggression and rule-breaking were related to informant, with higher heritability for aggression than rule-breaking only in the case of mother and teacher reports, and not for father or child reports. With regard to age, Burt (2009a) found that heritability for aggression and rule-breaking (averaged) ranged from $a^2 = .566$ for ages 1-5; $a^2 = .614$ for ages 6-10; and $a^2 = .547$ for ages 11-18. These findings were comparable to those obtained by Burt (2009b). Findings from both of the meta-analyses by Burt (2009a; 2009b) led us to explore the predictive distribution of heritability, given both informant and age, while controlling for all other moderators in our regression model. Furthermore, we sought to do so with data from longitudinal studies by informant and age, as Burt (2000a) had derived patterns in heritability by age largely from cross-sectional studies.

Bergen et al. (2007) conducted a meta-analysis with eight longitudinal studies measuring age-related changes in heritability of antisocial behavior over the course of adolescence and young adulthood. They employed a broad definition of antisocial behavior, including aggression, CD, Oppositional Defiant Disorder (ODD), and early problem behavior, and included heritability estimates from twin and adoption studies (both cross-sectional and longitudinal) at two or more time points. They excluded longitudinal studies using different informants over time, and controlled for gender in linear regression models; Bergen et al. (2007) found that heritability of externalizing behavior increased from age 10 ($h^2 = .28$) to age 15 ($h^2 = .30$) to age twenty ($h^2 = .35$). Although we were led by Bergen et al. (2007) to conduct a study of heritability over time, our approach was slightly different. First, we focused solely on antisocial behavior, and sought to include those studies employing different informants at different ages, controlling for that in our statistical model. Second, we sought to include measures of ethnicity, SES, and study location as moderators, as diversity of among samples of twin and non-twin siblings has increased. Third, we sought to explicitly address the issue of informant as a moderator, and focus on the ages of early childhood through adolescence in longitudinal research.

**Purpose of the Study**

The purpose of the present study was to conduct a systematic quantitative review of cross-sectional and longitudinal research in the behavioral genetics of antisocial behavior using multiple informants with children and adolescents. The goal of the meta-analysis was to examine patterns in the distribution of heritability across studies, identify significant moderators, and determine the extent to which heritability varied as a function of age and informant, particularly in longitudinal research. This study is important because the assessment of child and adolescent mental health is dependent upon informants' ratings, and work in advanced behavioral genetics depends upon informants' ratings. Thus, our study contributes to two substantive areas: (a) it builds upon previous meta-analytic research in the heritability of antisocial behavior, exploring longitudinal associations (Bergen et al., 2007; Burt, 2009a, 2009b; Miles & Carey, 1997; Rhee & Waldman, 2002) and (b) it contributes to ongoing research in the field of informant discrepancies (Achenbach, 2011; De Los Reyes et al., 2009; De Los Reyes & Kazdin, 2005; Wakschlag et al., 2008).

We sought to be comprehensive in our inclusion of studies for this analysis. First, we sought data from independent samples in longitudinal and cross-sectional research using multiple informants. Second, we employed a broad definition of antisocial behavior, including the perspectives of psychiatry, developmental psychopathology, and delinquency/rule-breaking behavior (Loeber, Burke, & Pardini, 2009; Rhee & Waldman, 2002). Third, we included only independent samples in the analysis and employed direct analysis of the data. Fourth, we examined the same study characteristics as moderators in our meta-analysis as did previous meta-analyses (Bergen et al., 2007; Burt, 2009a, 2009b; Miles & Carey, 1997; Rhee & Waldman, 2002). These included (a) participant characteristics: i.e., number of pairs and type of sibling relationship (MZ twin, DZ twin, biological full sibling, half sibling, unrelated sibling), mean age of pairs, gender, race, socioeconomic status (SES), zygosity determination method; (b) antisocial construct (aggression, delinquency, externalizing); (c) type of assessment (diagnostic interview or questionnaire); (d) informant characteristics (mother, father, teacher, observer, self);

and (e) study method characteristics (i.e., whether the sample was representative of the population; whether the study was cross-sectional or longitudinal; study wave and location).

## Method

### Procedures for Study Identification

Our search for studies was a three step process. First, we conducted an electronic search for abstracts from the earliest years through 2011 using the following databases: Google Scholar, Medline, PsychEXTRA, PsychInfo, and PubMED for studies using the following terms: antisocial, aggression, aggressive, behavior problems, conduct, delinquency, delinquent, externalizing, paired with each of the following search terms (*adoptee, adoptees, adoptive, behavioral genetic, twin, twins, environment*) (Burt, 2009b). Use of initial search terms in these databases resulted in 1,260 citations identified. We also reviewed studies that had been included by Rhee and Waldman (2002) and Burt (2009a, 2009b) and conducted additional ancestral searches of the reference lists of studies. Results from electronic and ancestral searches often overlapped; ten additional studies were identified by ancestral searches, bringing the total number of studies reviewed at step one to 1,270.

Studies were excluded at step one if they did not meet the broad criteria of being empirical investigations of antisocial behavior (using the psychiatric, developmental psychology, or delinquency/rule-breaking definitions) among children and youth (birth to age 21) using a twin, adoption, or sibling research design, with correlations between sibling pairs as effect sizes. We elected not to include studies containing correlations between parents and children (Burt 2009b). The following are examples of studies that were excluded following the electronic search: studies of adults; studies of alcohol, drug use, smoking, addiction, and suicide; studies of child adjustment and negative emotions; studies of depression and internalizing disorders;

studies from books and book chapters (unless original data were not available in journal articles); meta-analyses or literature reviews; and non-behavioral genetic studies. Studies of gene-environment interplay (i.e., variations in heritability according to environmental circumstances, correlations between genes and environments, interactions between specific identified genes and specific measured environments) were also excluded at step one (Rhee & Waldman, 2002).

Following the electronic and ancestral search processes, 989 studies were excluded and 281 studies fit the broad criteria for inclusion at step one. In a second step, this pool of studies was evaluated for construct validity and availability of correlations, which resulted in the exclusion of 144 additional studies. In a third and final step, the remaining 137 studies were reviewed for independence and nonindependence of samples, with selection favoring longitudinal research and multiple informants, after which 60 studies remained. We were able to obtain or calculate heritability statistics from 42 of the 60 studies. Appendix 1 contains a table (available from the study authors) listing all 137 studies, along with characteristics of samples assessed in each (age, gender, relationship between pairs of siblings, sample size); study characteristics (antisocial construct, informant, whether the study was longitudinal or cross-sectional and comprised a representative sample of the population); and effect sizes (as well as whether or not correlations were included in the meta-analysis).

**Inclusion Criteria for the Final Pool of Studies**

**Construct validity.** In the second step of the review process (which resulted in the exclusion of 144 studies), studies were included if they measured the construct of antisocial behavior from the perspectives of psychiatry, developmental psychopathology, and criminology. From psychiatry, this definition centers on diagnoses of CD and ASPD from the *Diagnostic and Statistical Manual of Mental Disorders* (4th Ed., *DSM-IV*); from developmental psychology, the definition includes the perspective of individual differences in aggressive and/or externalizing behavior as they unfold across development; and from criminology, the definition focuses on delinquent acts or and/or rule-breaking behavior (Loeber et al., 2009).

      *Psychiatric diagnoses*. In the *DSM-IV*, CD is described as "a repetitive and persistent pattern of behavior in which the basic rights of others or major age-appropriate societal norms or rules are violated," specifically aggression to people and animals, destruction of property, deceitfulness or theft, and serious violations of rules (American Psychiatric Association, 1994, p. 90). The onset of CD is during childhood or early adolescence (Rhee & Waldman, 2002). Following a history of CD before the age of 15, a diagnosis of ASPD may be given at age 18. ASPD includes "a pervasive pattern of disregard for and violation of the rights of others occurring since age 15 years," (American Psychiatric Association, 1994, p. 649), with the following characteristics: failure to conform to social norms with respect to lawful behaviors, deceitfulness, impulsivity, irritability and aggressiveness, reckless disregard for safety of self or others, consistent irresponsibility, and lack of remorse (American Psychiatric Association, 1994).

      Per Rhee & Waldman (2002), we excluded studies of personality associated with antisocial behavior. Forty-two studies of callous-unemotional personality traits were excluded, as were eight studies of related personality traits (i.e., interpersonal style of glibness, grandiosity, and manipulation and behavioral lifestyle of impulsivity and irresponsibility) (Larsson, Andershed, & Lichtenstein, 2006). However, measures of psychopathy associated with the construct of delinquency and collected from earlier studies were included (Loehlin & Nichols, 1976). This was because the latter was synonymous with ASPD using the Psychopathic Deviate subscale of the *Minnesota Multiphasic Personality Inventory*, (*MMPI*; Hathaway & McKinley, 1942), or the Socialization scale of the *California Psychological Inventory* (Gough, 1957).

In the review of studies of psychiatric diagnoses, correlations between siblings on measures of Attention Deficit Hyperactivity Disorder (ADHD) and Oppositional Defiant Disorder (ODD), 79 studies, were excluded. Some studies included measures of ADHD and ODD as well as CD; in these cases, measures of ODD were excluded during the evaluation of independence/nonindependence of samples at step three, and excluded from the table of studies included in Appendix 1.

*Aggression and externalizing behavior.* As noted by Rhee and Waldman (2002), the operationalization of aggression in the past has been very heterogeneous, ranging from personality traits to number of hits to a Bobo doll. For the present study, we included studies of aggression that reflected behavioral criteria in the *DSM-III-R* or the *DSM-IV* for CD (i.e., bullying and physical fighting), as well as items such as that were identical to those from the aggressive behavior scale of the Achenbach family of measures: arguing, bragging, being mean, demanding attention, destroying others' things, creating disturbances at home and at school, fighting, attacking, screaming, showing off, having temper tantrums, teasing, threatening, and being loud (Achenbach & Rescorla, 2001). Studies of externalizing behavior, a term which was coined by Achenbach and Edelbrock (1978) to describe problem behavior encompassing both aggression and delinquency (or rule-breaking behavior), were also included in the present study. In addition, seven behavioral genetic studies of the related construct of social or relational aggression were excluded.

*Delinquency/rule-breaking behavior.* According to Rhee & Waldman (2002), criminality describes an unlawful act that leads to arrest, conviction, or incarceration, and delinquency describes as unlawful acts committed by juveniles. Burt (2009a), per Achenbach and Rescorla (2001) uses the term rule-breaking instead of delinquent behavior to refer to covert or concealed antisocial behaviors, such as stealing, lying, drinking, destroying property, and burglary (Loeber & Schmaling, 1985).

**Inability to calculate correlations.** Estimates of heritability were the effect size(s) selected for the present study. These estimates were derived from intraclass Pearson product-moment correlations between MZ and DZ twin pairs (Burt, 2009a, 2009b; Rhee & Waldman, 2002), or retrieved from original research. In the case of seven studies, correlations were not reported and could not be calculated, so these studies were excluded.

**Assessment of Samples for Independence/Nonindependence**

In a third step, we reviewed 137 studies for final inclusion or exclusion on the basis of independence or nonindependence of study samples. We identified three types of cases in which independence/nonindependence was an issue. In the first case, the same data had been published in more than one study. In the second case, authors of a single publication included more than one measure of antisocial behavior in the study, which was typically associated with different informants. In the third case, with longitudinal research, authors obtained data from the same sample of children and youth over the course of time. We managed these three cases in the following ways.

In the first case, if the same data had been published in more than one study, we selected data to include from just one of the studies, applying the criteria for our research. That is, we selected those data for inclusion that had reported the greatest variety of informants. For example, if one study author had published results using both teacher and parent ratings, and a second study author published results using the same data, but with parent ratings only, then we selected studies for inclusion that had published results using both teacher and parent ratings.

This practice resulted in the emergence of the second case: the inclusion of multiple dependent variables associated with a single study, representing data from multiple informants or assessments of the same participants at a different age/wave. To manage this in the analysis, we included type of informant as a predictor in our regression model, so that studies employing more than one informant would not have a greater effect on the dependent variable. In addition, in our initial analysis, we allowed the regression model to check for predictive differences between the different informants.

In the third case, in keeping with both the purpose of our research and seeking to maintain independence of samples, we sought to include data from longitudinal studies with results for the same sample of children and youth collected over time, at different waves. We selected independent samples representing multiple waves of data in as many cases as possible, including those with the greatest variety of informants. We then managed the waves of data by identifying each study in the meta-analysis as a member (or not) of a particular longitudinal data set, associated with a particular wave of data collection by mean age of the study sample. So, for example, using the E-Risk longitudinal data set, we included four studies (of eight possible) in our analyses. Data were included from mothers' ratings of children at age three, wave one (Gregory, Eley, & Plomin, 2004); from mother, observer, teacher, and self-ratings of the same children at age five, wave two (Arseneault, Moffitt, Caspi, Taylor, Rijsdijk, & Jaffee, 2003); from mother and teacher ratings of the same children at age seven, wave three (Saudino, Ronald, & Plomin, 2005); and from mother and teacher ratings at age 10, wave four (Ball, Arsenault, Taylor, Maughan, Caspi, & Moffitt, 2008). In some of these cases, data had already been combined across informants; in other cases, they had not been. Our data entry reflected that, along with the age and wave associated with each longitudinal sample.

**Inability to Calculate or Retrieve Heritability**

Heritability was calculated using the Falconer and MacKay (1996) estimator of heritability, $h^2$ ($n = 29$ studies) or retrieved directly from the research ($n = 13$ studies) (in the next section, we provide more technical details about the estimator). In the latter case, we recorded the heritability statistic ($h^2$ or $a^2$), and the corresponding sampling variance associated with the statistic. The majority of these studies ($n = 9/13$ studies, 69%) reported heritability using behavioral genetic modeling procedures (along with confidence intervals) and controlling for gender.

## Study Coding Procedures

Data were coded to capture study characteristics expected to serve as potential moderators of heritability, based on previous meta-analyses. In coding, each of the characteristics of a given study was paired with a correlation between twins and/or non-twin siblings on a measure of antisocial behavior; these correlations were then used to calculate heritability. Study characteristics included (a) participant characteristics (sibling pairs), (b) antisocial construct and type of assessment (i.e., diagnostic interview or questionnaire), (c) informant characteristics, and (d) study method characteristics.

**Participant Characteristics**

Participant characteristics recorded were those associated with pairs of siblings, including number of pairs, type of sibling relationship (i.e., MZ twin, DZ twin, biological full sibling, half sibling, unrelated sibling) mean age of pairs, and gender of pairs (male, female, or opposite sex). In some cases, authors did not report gender by sibling pair, so we calculated percent of the participants in the sample who were MZ female, MZ male, DZ female, DZ male, etc. We also recorded percent of participants in the study from various racial backgrounds, including White,

Black, Latino/a, Asian, and other. Parent SES was coded according to data that authors provided, with included parent occupation, parent education, and/or income. Data for SES were recorded as low when the majority of the sample (60% or more) was comprised of individuals with parent education levels that were below $12^{th}$ grade level or who held jobs at low levels (for example, levels I-III) of the Hollingshead (Hollingshead & Redlich, 1958) or similar occupational ranking procedure. Data for SES were recorded as middle-to-high when the majority of the sample (60% or more) was comprised of individuals with parent education levels that were $12^{th}$ grade level and above, or who held jobs at the highest levels of the Hollingshead or similar occupational ranking procedure. Data for SES were recorded as missing when parent SES was not provided in any form (neither parent education nor occupation nor income). In cases where study authors reported a range of participants with regard to SES, data for SES were recorded in both categories (i.e., low and middle-to-high).

As noted previously, estimates of the magnitude of genetic and environmental influences may be affected by the method for assessing zygosity (Rhee & Waldman, 2002). In the present study, data for method of zygosity determination were recorded as follows: (a) researchers used a questionnaire to determine zygosity; (b) researchers used a collection of DNA samples (cheek swabs, blood samples, etc.) to determine zygosity; (c) researchers used both questionnaire and DNA methods to determine zygosity of sibling pairs; or (d) data were missing for method of zygosity determination.

**Antisocial Construct and Method of Assessment**

Diagnosis of psychiatric disorders (CD and ASPD) is frequently determined through clinical interviews of parents and youth using tools such as the *Diagnostic Interview Schedule for Children, DISC* (Shaffer, Fisher, Lucas, Dulcan, & Schwab-Stone, 2000). Data were recorded for CD if the study clearly described diagnoses or symptom counts associated with the *DSM-III* or *DSM-IV* criteria. Because Burt (2009b) had reported a significant difference between the use of diagnostic interviews and questionnaires in her model, we coded this information and included it as a moderator.

Prior to the final meta-analysis and due to their relatively low numbers, studies originally coded under the construct CD (i.e., Burt, McGue, Krueger, & Iacono, 2005; Burt, McGue, & Iacono, 2010) were collapsed with studies coded under aggressive behavior. Likewise, studies coded under the construct ASPD (i.e., Burt et al., 2010) were collapsed with studies coded under delinquency. However, the assessment of psychiatric diagnoses was retained in both of these cases, as we entered codes for diagnostic interview for these studies and their respective samples.

Aggression has been assessed using a variety of methods, including parent, teacher, and self reports, along with observational measures. In the present study, measures of aggression were obtained via the Achenbach family of instruments (Achenbach & Rescorla, 2001): *Child Behavior Checklist (CBCL)*, *Teachers Report Form* (*TRF)*, and *Youth Self-Report* (*YSR)*; the *Behavior Events Inventory* (Patterson, 1982), *Behavior Problem Index* (Zill, 1985), Olweus self-report questionnaire (Olweus, 1989), and *Strengths and Difficulties Questionnaire* (Goodman, 1997). In each of these questionnaires, data for aggression were obtained from symptom counts of overt behaviors such as frequent arguing, being mean, destroying things, being disobedient at home and at school, getting in fights, attacking people, screaming a lot, having temper tantrums, etc. (We excluded, as did Rhee & Waldman (2002) and Burt (2009b) studies of aggressive behavior measured by number of hits to a Bobo doll.)

The Achenbach family of instruments has also been employed to assess rule-breaking and delinquent behavior (Achenbach & Rescorla, 2001); in addition, researchers in the present study employed *Delinquent Behavior Among Young People in the Western World* (Junger-Tas, Terlouw, & Klein, 1994). Assessment of the construct delinquency/rule-breaking includes symptom counts of rule-breaking and illegal behavior for minors, such as drinking alcohol, not feeling guilty after misbehaving, breaking rules, vandalizing property, hanging around with others who get in trouble, lying, cheating, stealing, etc. Data for the construct externalizing included symptom counts of both aggression and delinquency, typically combined in a common measure (Achenbach & Edelbrock, 1978). Because one third of the 42 studies ($n = 14$) employed the Achenbach family of measures, we coded this information and included it as a moderator as well.

In addition to diagnostic interviews and responses to questionnaires, observations of aggressive and antisocial behavior have also been employed by researchers. For example, in the case of the E-risk study observers completed the *Dunedin Behavioural Observation Scale* (Caspi, Henry, McGee, Moffitt, & Silva, 1995) following a home visit, and in the case of the Nonshared Environment Adolescent Development study (NEAD), observers coded dyadic videotaped interactions between each adolescent and his or her mother, father, and sibling (Hetherington & Clingempeel, 1992).

**Informant Characteristics**

Data were recorded for each informant associated with a given measure of antisocial behavior. First, we recorded data when single informants were employed; we also recorded data for multiple informants, among them mothers, fathers, teachers, self and/or observers. In some cases, different informants were associated with different measures reported in a given study; in other cases, informants were combined prior to the authors' calculation of correlations. Data were coded to reflect both of those cases. When data for informants was combined by authors, we recorded the proportion of study participants who were associated with a particular informant.

**Characteristics of Study Methods**

Studies were coded according to whether they employed a twin, non-twin sibling, or adoption research design. To capture this, we first coded the type of relationship between siblings that was associated with a given correlation: MZ twin, DZ twin, non-twin sibling (and whether the sibling relationship was full biological, half, or no biological relationship). Each of the types of the sibling pairs was in turn tied to the type of study: twin, adoption, or sibling study (non-twin or adoptive).

In order to capture additional important study method characteristics for the meta-analysis, we coded whether a study employed a longitudinal (including wave of data) or cross-sectional research design. In addition, we coded whether the study sample was a representative or a convenience sample, and the study location with regard to longitude and latitude.

**Evaluation of Coding Procedures**

Two independent raters coded 19 of the included studies (45%). These studies had been randomly selected for independent coding in each of the following categories/study characteristics: (a) participant characteristics (sibling pairs), (b) antisocial construct and type of assessment (i.e., diagnostic interview or questionnaire), (c) informant characteristics, and (d) study method characteristics. Following the development of a coding manual and a training period, we conducted agreement checks between coders and identified problem codes, which were then redefined. Studies were then recoded under refined definitions, and new studies recoded. The two independent raters obtained 100% agreement across the 58 codes representing 29 moderators/moderator variables in 19 studies.

**Heritability as Effect Size for the Meta-analysis**

We selected heritability as our effect size, which was derived from correlations in antisocial behavior among pairs of siblings, with antisocial behavior assessed by behavioral ratings made by parents, teachers, observers, and/or youth. Overall, we obtained $n = 89$ heritability estimates from 94,517 sibling pairs in 42 studies (38 were studies of twins; 3 were studies of twins and non-twin siblings; 1 was an adoption study). For 29 of the 42 studies (69%), representing 71 heritability estimates, we calculated heritability from correlations between MZ twins and their same gender DZ twin counterparts (in two different ways, depending upon number of informants in the study, which we describe below). For 13 of the 42 studies (31%) representing 18 samples of sibling pairs, we retrieved heritability statistics (and their corresponding sampling variances) from original research. In the majority of the 13 studies ($n = 9$, 69%), heritability had been estimated via genetic modeling procedures, controlling for gender and other study characteristics. To account for gender in these 13 studies, we recorded the percentage of males and females that had been used to obtain these model-based heritability estimates.

We had nearly twice as many estimates of heritability ($n = 89$) as studies ($n = 42$) for the following reasons. First, 25 of the 42 studies (62%) resulted in more than one estimate because we calculated or recorded separate heritability statistics for male and female sibling pairs. Second, 2 of the 42 studies (5%) resulted in more than one estimate because studies assessed sibling pairs at different ages. Two additional studies of the 42 (5%) resulted in more than one estimate because heritability statistics were listed separately by informant. One of 42 studies (2%) resulted in more than one estimate because heritability was listed separately by construct. Finally, in the case of 12 studies (29%), we obtained one heritability statistic per study.

In the case of 29 of the 42 studies (69%) representing 71 samples of sibling pairs, we calculated heritability from correlations between MZ twins and their same gender DZ twin counterparts in the following two ways. First, for a given gender, suppose that $n_{MZ}$ MZ twins yield a sample correlation $\hat{\rho}_{MZ}$ on an antisocial construct, and that $n_{DZ}$ DZ twin pairs yield a correlation $\hat{\rho}_{DZ}$ on the same construct. Then an estimate of the heritability of antisocial behavior is given by:

$$\hat{h}^2 = 2(\hat{\rho}_{DZ} - \hat{\rho}_{DZ}) \qquad (1)$$

(Falconer & Mackay, 1996). It can be shown that this estimate has sampling variance:

$$\hat{\sigma}^2 = 4[\{(1-\rho_{MZ}^2)^2/n_{MZ}\} + \{(1-\rho_{DZ}^2)^2/n_{DZ}\}]. \qquad (2)$$

We used this method for $n = 17$ of the 42 studies (40%) and their respective samples of sibling pairs ($n = 51$; 57% of sibling pairs), which included data for one informant. For $n = 12$ of the 42 studies (29%) and their respective samples of sibling pairs ($n = 20$; 22% of sibling pairs), we had data for different types of informants rating the same set of siblings. Therefore, in the case of the 12 studies, we obtained a within-study inverse-variance weighted average of heritability estimates. This facilitated a more interpretable univariate meta-analysis across all samples of sibling pairs. Finally, for $n = 13$ of the 42 studies (31%) and their respective samples of sibling pairs ($n = 18$; 20% of sibling pairs), we obtained heritability estimates (and their corresponding sampling variances) directly from the original studies,

Figure 1 presents a plot the heritability estimates and corresponding sampling variances for each of the $n = 89$ heritability estimates. Table 1 presents univariate descriptive statistics for the $n = 89$ heritability estimates ($\hat{h}_i$, $i = 1,…,n$), and for their corresponding variances ($\hat{\sigma}_i^{-2}$, $i = 1,…,n$).

**Summary of Participant Characteristics**

With regard to gender, $n = 36$ (40%) of the 89 heritability estimates were obtained from female pairs; $n = 37$ (42%) were obtained from male pairs; and $n = 16$ (18%) were obtained from genetic models controlling for gender. With regard to SES status of participants, $n = 61/89$ (68%) heritability estimates were associated with sibling pairs from middle-to-high socioeconomic backgrounds; $n = 21$ (24%) were from low and middle-to-high socioeconomic groups (i.e., sibling pairs represented a range of SES); and $n = 7$ (8%) reported no socioeconomic characteristics for their samples. With regard to race of participants, for $n = 80$ of the 89 samples (90%), sibling pairs were at least 60% White. With regard to assessment of zygosity, $n = 35$ of the 89 samples (39%) were associated with either a questionnaire or a DNA assessment, whereas $n = 30$ (34%) were associated with a questionnaire only, and $n = 13$ (15%) were associated with a DNA assessment only. Just one study (1%), the single adoption study in our sample, did not assess zygosity (Burt et al., 2010).

**Summary of Antisocial Construct, Type of Assessment, and Informant Characteristics**

With regard to antisocial construct, $n = 44$ (50%) of the 89 heritability estimates were associated with externalizing problems, whereas $n = 36$ (40%) were associated with aggression and CD; and $n = 9$ (10%) were associated with delinquency. To obtain these estimates, the majority of researchers in the present study employed questionnaires ($n = 83$ of 89 heritability estimates, 93%), whereas $n = 6$ heritability estimates (7%) were associated with diagnostic interviews. Of the $n = 83$ heritability estimates that had been associated with questionnaires, $n = 44$ (53%) were associated with one or more of the Achenbach measures (Achenbach & Rescorla, 2001).

**Summary of Study Method Characteristics**

Data included in the present study were from cross-sectional studies and single waves of longitudinal research ($n = 17$ studies; 40%), as well as independent samples from multiple waves of longitudinal research ($n = 25$ studies; 60%). We found that $n = 72/89$ (81%) of the heritability estimates employed a representative sample of the population from which they drew, and $n = 17$ (19%) employed a convenience sample. Furthermore, $n = 72$ (81%) of the $n = 89$ heritability estimates were from longitudinal research. As the table in Appendix 1 indicates, we frequently included multiple independent studies within a longitudinal set of data. For the following longitudinal data sets, we included two or more waves of data: Environmental Risk Longitudinal Twin Study (E-Risk), Minnesota Twin Family Study, Quebec Newborn Twin Study, Oregon Twins, Sibling Interaction and Behavior Study (SIBS), Dutch Twins, Swedish Twin Study (TCHAD), Colorado Twin Registry, NEAD, University of Southern California (USC) Twin Project, and Finish Twins (Finn Twin).

Figure 1 reveals that the majority of sibling pairs were rated by mothers ($n = 34/89$ samples of sibling pairs, 38%), followed by ratings from mixed informants ($n = 24/89$, 27%), teacher informants ($n = 15/89$, 16%), observer informants ($n = 3/89$, 4%), and youth informants describing themselves ($n = 13/89$, 15%). Researchers collecting data from mixed informants describing $n = 24$ samples of sibling pairs used different combinations. The most frequent combinations were (a) some combination of a parent, either mother or father or both, with a teacher ($n = 10/24$, 42%); followed by (b) mother and father ($n = 6/24$, 25%); (c) mother and self ($n = 3$, 13%); and mother, father, teacher, and self ($n = 2/24$, 8%); and mother, teacher, observer, and self ($n = 2/24$, 8%).

The data for participants, antisocial construct, type of assessment, informant, and study method characteristics are summarized in this section and in the two prior sections. Together, these data are described as 29 moderator variables, corresponding to each of the $n = 89$ heritability estimates. Table 1 also presents univariate descriptive statistics for each of these moderators over the 89 cases.

**Meta-Analysis Methods**

For the meta-analysis, the 89 heritability estimates (effect sizes) were treated as observations of a dependent variable, with each estimate (observation) having a weight defined by the inverse of the sampling variance of the estimate (i.e., $\hat{\sigma}_i^{-2}$, $i = 1,\ldots,n$). Also, the 29 variables that describe study characteristics (see Table 1) were treated as moderators (predictors). In terms of notation, the full set of data is denoted by $\mathcal{D}_{89} = \{(y_i = \hat{h}_i^2, \hat{\sigma}_i^{-2}, \mathbf{x}_i)\}_{i=1}^{n=89}$, with moderators $\mathbf{x} = (1, x_1, \ldots, x_{29})^T$, including a constant (1) term for future notational convenience.

In the meta-analysis of our data, we had three goals. The first goal was to perform statistical inference of the "overall" heritability distribution, in the underlying population of studies. The second goal was to identify significant study-level moderators (predictors) of heritability (or changes in the mean heritability). The third goal was to closely examine how the heritability distribution varied as a function of child age and informant type, particularly in longitudinal research.

Given the prior research, our belief is that the distribution of child and adolescent psychopathology in the population is likely to be positively skewed, with relatively few individuals in the population experiencing disorders (Burt, 2009b). Thus, it seemed reasonable to assume that the distribution of the heritability estimates would not be adequately described by a simple unimodal and symmetric distribution, such as a normal distribution. Moreover, it was reasonable to believe that the entire heritability distribution may not necessarily depend linearly on the 29 moderators.

Given all the above considerations, we performed a meta-analysis of the data using a flexible, Bayesian nonparametric random-effects regression model, which was first introduced by Karabatsos and Walker (2012). This is an infinite-mixture model which allows the entire distribution of the dependent variable to change flexibly and nonlinearly (or linearly) as a function of moderators. Specifically, the model is defined by an infinite-mixture of normal densities (distributions), with mixture weights that depend on the moderators (or predictors). The development of this model was motivated by the well known statistical result, that any smooth probability distribution (density) could be approximated arbitrarily well by a suitable mixture of normal distributions (densities). The Bayesian model is "nonparametric" in the sense that it does not strictly assume that a data distribution can only be described by a small number of parameters. Instead, such a model can describe a large range of (smooth) data distributions, from simple unimodal symmetric distributions that are fully-describable by a small number of parameters, to more skewed and multimodal distributions which are more adequately described by very many or infinitely many parameters (Müller & Quintana, 2004). In contrast, models that assume the normal distribution assume that the data distribution can only be unimodal and symmetric, and can be adequately described by a small number of parameters, such as the mean and variance.

Karabatsos and Walker (2012) demonstrated that their Bayesian nonparametric model tended to have better predictive performance than many other regression models of common usage, for a wide range of real data sets, and for data sets simulated under a wide range of complex data generation models. Later, for more specialized meta-analytic settings, Karabatsos,

Walker, & Talbott (2012) demonstrated that the Bayesian model had good predictive performance for a range of meta-analytic data sets. In other words, for each meta-analytic data set, the Bayesian model tended to describe well the effect size (e.g., heritability) distribution for the underlying population of studies.

Mathematically, for the data set $\mathcal{D}_{89} = \{(y_i = \hat{h}_i^2, \hat{\sigma}_i^{-2}, \mathbf{x}_i)\}_{i=1}^{n=89}$ of the current study, the Bayesian nonparametric model (Karabatsos & Walker, 2012) specifies the probability density $f(y \mid \mathbf{x})$ (distribution) of the heritability $y$, conditional on any vector of 29 moderators $\mathbf{x} = (1, x_1, \ldots, x_{29})^T$, according to the infinite-mixture model:

$$\begin{aligned} f(y_i \mid \mathbf{x}_i; \zeta) &= \int \mathrm{n}(y_i \mid \beta_0 + \beta_1 x_{1i} + \ldots + \beta_{29} x_{29,i} + \mu, \phi\hat{\sigma}_i^2) \mathrm{d}G_{\mathbf{x}_i}(\mu) \\ &= \sum_{j=1}^{\infty} \mathrm{n}(y_i \mid \beta_0 + \beta_1 x_{1i} + \ldots + \beta_{29} x_{29,i} + \mu_j, \phi\hat{\sigma}_i^2)\, \omega_j(\mathbf{x}_i; \boldsymbol{\beta}_\omega, \sigma_\omega), \quad i = 1,\ldots, n = 89. \end{aligned} \quad (3)$$

Above, $\zeta = (\boldsymbol{\mu}, \boldsymbol{\beta}, \phi, \sigma_\mu^2, \boldsymbol{\beta}_\omega, \sigma_\omega)$ denote the model parameters, including the random-effect parameters $\boldsymbol{\mu} = (\mu_j \mid j = 0, +1, +2, \ldots)$, the linear regression slope coefficients $\boldsymbol{\beta} = (\beta_0, \beta_1, \ldots, \beta_p)$, the dispersion parameter $\phi$ (Thompson & Sharp, 1999), the variance parameter $\sigma_\mu^2$ of the random effects, along with parameters $(\boldsymbol{\beta}_\omega, \sigma_\omega)$ of the mixture weights which will be described later. Also, $\mathrm{n}(\cdot \mid \mu, \sigma^2)$ denotes the probability density function for the normal distribution having mean and variance $(\mu, \sigma^2)$; that is, it is the function that defines the "bell-shaped curve." As shown in our model (3), the probability density $f(y \mid \mathbf{x})$ (distribution) of heritability, given any specific moderators $\mathbf{x}$ of interest, is formed by an infinite mixture of normal densities (distributions), with random effect (intercept) parameters $\mu_j$ (for $j = 0, \pm 1, \pm 2, \ldots$), and with mixture weights $\omega_j(\mathbf{x}_i; \boldsymbol{\beta}_\omega, \sigma_\omega)$ for these random effects (intercepts) that depend on the moderators (predictors), and correspond to the mixing distribution $G_\mathbf{x}(\mu)$. Furthermore, the mixture weights $\omega_j(\mathbf{x}_i; \boldsymbol{\beta}_\omega, \sigma_\omega)$ sum to one for each distinct value of the moderators $\mathbf{x}$. These weights are defined by an ordered-probits regression, for the ordinal categories $j = 0, \pm 1, \pm 2, \ldots$, according to:

$$\omega_j(\mathbf{x}_i; \boldsymbol{\beta}_\omega, \sigma_\omega^2) = \Phi(\{j - \mathbf{x}^T\boldsymbol{\beta}_\omega\}/\sigma_\omega) - \Phi(\{j - 1 - \mathbf{x}^T\boldsymbol{\beta}_\omega\}/\sigma_\omega) \qquad (j = 0, \pm 1, \pm 2, \ldots), \quad (4)$$

where $\Phi(\cdot)$ denotes the standard Normal(0,1) cumulative distribution function (c.d.f.), where $\mathbf{x}^T\boldsymbol{\beta}_\omega = \beta_{\omega 0} + \beta_{\omega 1} x_1 + \ldots + \beta_{\omega 29} x_{29}$, with $\boldsymbol{\beta}_\omega = (\beta_{\omega 0}, \beta_{\omega 1}, \ldots, \beta_{\omega 29})$ the linear slope coefficients of the ordered-probits regression, and where $\sigma_\omega$ the error standard deviation of the probit regression. In summary, our Bayesian nonparametric model (3) can thus be viewed as a type of random effects model for meta-analysis, where the random intercepts distribution $G_\mathbf{x}(\mu)$ (and therefore, the effect size density $f(y|\mathbf{x})$) can change flexibly and nonlinearly as a function of the moderators $\mathbf{x}$.

The Bayesian model (3) is completed by the specification of prior densities (distributions), namely $\mu_j \sim_{\text{i.i.d.}} n(0, \sigma_\mu^2)$, $\beta_0 \sim n(0, 10^5)$, $\beta_k \sim n(0,1)$ for $k = 1, \ldots, p = 29$, $\phi \sim \text{ga}(1/2, 1/2)$, $\sigma_\mu^2 \sim u(0,100)$, $\beta_{\omega k} \sim n(0, 10^5)$ for $k = 0, 1, \ldots, p = 29$, and $\sigma_\omega^{-2} \sim \text{ga}(1, 1)$. Here, $\sim$ means "distributed as", $\sim_{\text{i.i.d.}}$ means "independently and identically distributed, while $u(\cdot|0,100)$ denotes a probability density of a uniform distribution, and $\text{ga}(\cdot|a,b)$ denotes the density of the gamma distribution with shape and rate parameters $(a,b)$. These prior distributions have relatively high prior variance, but adequately represent our actual prior beliefs about the parameters. Moreover, the priors we assigned for the regression parameter $\beta_0$, and for the dispersion parameter $\phi$, are consistent with the general recommendations made for Bayesian meta-analytic models (DuMouchel & Normand, 2000; Nam, Mengersen, & Garthwaite, 2003), and the prior we assigned to the random-effects variance parameter $\sigma_\mu^2$ is consistent with a general recommendation made for Bayesian random-intercepts models (Gelman, 2006). The priors we assigned to the slope parameters $\beta_k$, $k = 1,\ldots,p = 29$ reflected the size of the scale of the heritability measure, which ranged in the interval $(0,1)$. Finally, the priors that we specified for the mixture weight parameters $(\boldsymbol{\beta}_\omega, \sigma_\omega^2)$ seemed to be reasonable for an earlier study involving a more general version of the Bayesian nonparametric model (Karabatsos & Walker, 2012). The full joint prior density of all the model parameters is expressed by

$$\pi(\boldsymbol{\mu}, \boldsymbol{\beta}, \phi, \sigma_\mu^2, \boldsymbol{\beta}_\omega, \sigma_\omega^2) = \left\{\prod_j \mathrm{n}(\mu_j \mid 0, 1/2)\right\} \mathrm{n}(\beta_0 \mid 0, 10^5) \left\{\prod_{k=1}^{29} \mathrm{n}(\beta_k \mid 0, 1)\right\} \mathrm{ga}(\phi \mid 1/2, 1/2) \\ \times \mathrm{un}(\sigma_\mu^2 \mid 0, 100) \left\{\prod_{k=0}^{29} \mathrm{n}(\beta_{\omega k} \mid 0, 10^5)\right\} \mathrm{ga}(\sigma_\omega^{-2} \mid 1, 1). \tag{5}$$

According to standard arguments involving Bayes' theorem, the full data set $\mathcal{D}_{89} = \{(y_i = \hat{h}_i^2, \hat{\sigma}_i^{-2}, \mathbf{x}_i)\}_{i=1}^{n=89}$ combines with the prior density function $\pi(\boldsymbol{\mu}, \boldsymbol{\beta}, \phi, \sigma_\mu^2, \boldsymbol{\beta}_\omega, \sigma_\omega^2)$, to yield a posterior density function, which is given by

$$\pi(\boldsymbol{\mu}, \boldsymbol{\beta}, \phi, \sigma_\mu^2, \boldsymbol{\beta}_\omega, \sigma_\omega^2 \mid \mathcal{D}_{89}) \propto \prod_{i=1}^{89} f(y_i \mid \mathbf{x}_i; \boldsymbol{\zeta}) \pi(\boldsymbol{\mu}, \boldsymbol{\beta}, \phi, \sigma_\mu^2, \boldsymbol{\beta}_\omega, \sigma_\omega^2), \tag{6}$$

up to a proportionality constant. In words the posterior density (distribution) gives the plausible values of all the model's parameters, given the data $\mathcal{D}_{89}$ having likelihood $\prod_{i=1}^{89} f(y_i \mid \mathbf{x}_i; \boldsymbol{\zeta})$ under the Bayesian model (3), and given the prior $\pi(\boldsymbol{\mu}, \boldsymbol{\beta}, \phi, \sigma_\mu^2, \boldsymbol{\beta}_\omega, \sigma_\omega^2)$. Also, for the purposes of predicting plausible values of the heritability variable $Y$, conditional on the moderator values $\mathbf{x}$ of interest, we have the posterior predictive distribution of heritability given $\mathbf{x}$, which is defined by the probability density:

$$f_n(y \mid \mathbf{x}) = \int f(y \mid \mathbf{x}; \boldsymbol{\zeta}) \pi(\boldsymbol{\mu}, \boldsymbol{\beta}, \phi, \sigma_\mu^2, \boldsymbol{\beta}_\omega, \sigma_\omega^2 \mid \mathcal{D}_{89}) \mathrm{d}\boldsymbol{\mu}\, \mathrm{d}\boldsymbol{\beta}\, \mathrm{d}\phi\, \mathrm{d}\sigma_\mu^2\, \mathrm{d}\boldsymbol{\beta}_\omega \mathrm{d}\sigma_\omega^2. \tag{7}$$

The posterior predictive density provides a sample estimate of the heritability distribution for the underlying population studies, conditional on any moderators $\mathbf{x}$ of interest (Aitchison, 1975). The posterior probability densities (distributions) in (6) and (7) can be estimated using standard methods of Markov chain Monte Carlo (MCMC), which are described in Appendix A of Karabatsos & Walker (2012). Finally, prior to applying the Bayesian nonparametric model for meta-analysis, we z-standardized the data for each of the 29 moderators sample mean 0 and variance 1. But all the results of the meta-analysis will be reported on the original scale of the moderators.

Now we explain how the three goals of our meta-analysis are addressed, with regard to parameters of interest in our Bayesian nonparametric random-effects model defined by equations (3)-(5), in terms of the posterior distribution of these model parameters (6), and in terms of the posterior predictive probability density of the model (7). For the first goal, the overall heritability distribution is inferred from the posterior predictive probability density $f_n(y \mid \mathbf{x})$ of that distribution, conditional on covariates $\mathbf{x} = (1,0,\ldots,0)$, while controlling for all 29 moderators by fixing their standardized values to zero. Multiple modes in this distribution (density) would indicate the presence of multiple latent clusters of heritability in the underlying study population. For the second goal, the linear impact of each moderator $x_k$, on the mean heritability, is measured by summaries (e.g., mean, standard deviation, 95% interval) of the marginal posterior distribution of the linear regression slope parameter $\beta_k$ (for $k = 1,\ldots,29$). In other words, for a given moderating variable $X_k$, the slope $\beta_k$ indicates the change in the mean heritability, for every unit increase in the moderator, after controlling for all random effects (i.e., the $\mu_j$, for $j = 0, \pm 1, \pm 2, \ldots$) by setting them to zero, and after controlling for all the 28 other moderators (predictors) by fixing their standardized values to zero. Importantly, recall that one of the moderators was defined by the standard error (SE), i.e., the square root of the heritability variance $\hat{\sigma}^2$. The slope coefficient for that moderator would indicate how much publication bias affected the results of a meta-analysis (Thompson & Sharp, 1999). Finally, the third goal can be addressed by examining how key features of the heritability distribution, such as the median and inter-quartile range (variance) of the posterior predictive density $f_n(y \mid \mathbf{x})$ of that distribution, changes over a range of values of the child age moderator (age variable; see Table 1), while conditioning on moderators that reflect the given informant type of interest (i.e., conditional on the moderator of either Mom=1, Dad=1, teacher=1, Self=1, or observer=1, with all four of the other types of informant set to zero; see Table 1), conditioning on a moderator indicating longitudinal study (i.e., conditional on longsampl=1; see Table 1), while also conditioning all the other 23 moderators by fixing their values to zero.

In any regression analysis, including meta-analysis, it is important to evaluate the adequacy of the regression model for the given data set, in order to verify whether the model provides reasonably-accurate statistical inferences of the data. Therefore, for the heritability data set, we evaluated the predictive performance of our Bayesian nonparametric meta-analytic model, in order to evaluate how well the model described the heritability distribution for the underlying population studies, conditional on the observed values of the moderators $\mathbf{x}_i$, $i = 1,\ldots,n$. Specifically, we evaluated the model's performance through the use of a predictive mean-square error criterion that was introduced by Laud and Ibrahim (1995), and which was further studied by Gelfand and Ghosh (1998) from a Bayesian decision-theoretic perspective. For our Bayesian model, the predictive mean-square error criterion is written as:

$$D(m) = \sum_{i=1}^{n=89} \int (y_i - y)^2 f_n(y \mid \mathbf{x}) \mathrm{d}y = \sum_{i=1}^{n=89} \left\{ (y_i - \mathrm{E}_n(Y_i \mid \mathbf{x}_i))^2 + \mathrm{Var}_n(Y_i \mid \mathbf{x}_i) \right\} = \sum_{i=1}^{n=89} D_i(m). \qquad (8)$$

In the third term of (8), the quantities $\mathrm{E}(Y_i \mid \mathbf{x}_i)$ and $\mathrm{Var}(Y_i \mid \mathbf{x}_i)$ are, respectively, the expectation (mean) and variance from the posterior predictive probability density $f_n(y \mid \mathbf{x}_i)$, for $i = 1,\ldots,n$, under our Bayesian nonparametric model. Also in that third term, the square error term measures the predictive bias for the heritability data $y_i$ (for $i = 1,\ldots,n$), and the variance term $\mathrm{Var}(Y_i \mid \mathbf{x}_i)$ is a penalty term that is large if either the model over-fits or under-fits the data (Gelfand & Ghosh, 1998). Finally, in relation to the fourth term in (8), each quantity $D_i(m)$ gives the predictive mean-square error for the individual heritability observation $y_i$. Therefore, the individual square-root quantities $\sqrt{D_i(m)}$ (for $i = 1,\ldots,n$) can be used to provide a detailed assessment of the model's predictive performance on the original scale of the heritability $y_i$. A large value of $\sqrt{D_i(m)}$ would indicate that the heritability $y_i$ is an outlier under the model.

### Results of the Meta-Analysis

The Bayesian nonparametric meta-analysis model was estimated by 200,000 MCMC converged samples from the posterior distribution. Evidence of convergence was supported by trace plots, which presented adequate mixing of the MCMC samples of the model parameters, and was supported by small batch-means 95% Monte Carlo confidence intervals which have half-widths that typically were less than .05. These follow according to recommended procedures for evaluating the quality of posterior distribution estimates, on the basis of MCMC samples (Geyer, 2011).

For meta-analysis, our Bayesian nonparametric model adequately fit the heritability data. The MCMC estimate of the model's overall predictive mean square error $D(m)$ was .31. Also, the estimates of the individual square-root quantities $\sqrt{D_i(m)}$ had a 5-number summary (i.e., minimum, 25%ile, median, 75%ile, and maximum) values of .01, .02, .04, .05, and .25.

The top of Figure 2 presents the overall posterior predictive density estimate of the heritability distribution for the underlying population of studies. This estimate was obtained conditional on the effect size variance $\hat{\sigma}^2$ of .001, the minimum value found in the data set, in order to reflect information from a large sample study. The mean and median in this figure are .61, but there are also two modes in the distribution, namely at about .51 and .72, indicating two latent clusters of heritabilities in the study population. In addition, the distribution had more skewness (-.3) and less kurtosis (2.9) than a normal distribution. Moreover, the marginal posterior mean (S.D.) estimate of the dispersion parameter $\phi$ was .04 (.01), while the marginal posterior mean (S.D.) estimate for the random intercept variance $\sigma_\mu^2$ was also .04 (.01).

Table 2 presents the estimates of the marginal posterior mean and corresponding 95% credible (confidence) intervals. Each interval is formed by 2.5%ile and 97.5%iles of the marginal posterior of the given coefficient for each regression coefficient. The table indicates the moderators that serve as significant predictors of the mean heritability, after controlling for all 28 other moderators by setting them to zero. Such a significant moderator is indicated by a slope parameters ($\beta_k$) with marginal 95% posterior (confidence) interval estimate that does not include zero. As shown in this table, 23 of the 29 moderators did not significantly predict changes in mean heritability. The SE moderator appeared to be a marginally significant predictor of mean effect size (SE; slope estimate .09; 95% CI = (-.02, .17)). But we noticed in separate plots that the posterior predictive median of heritability did not significantly change as a function of SE, after taking into account the interquartile range. Hence, these results can be taken as evidence that publication bias does not strongly affect the results of the meta-analysis. Moreover, according to Table 2, heritability was on average significantly higher for studies that were based on an average of heritability measures versus not ($h^2$_Ave; slope estimate .05; 95% CI = (.01, .10)), was significantly higher among studies having at least 60% white twins and non-twin siblins versus not (White60; slope estimate .12; 95% CI = (.06, .18)), was significantly higher for longitudinal versus non-longitudinal studies (long; slope estimate .09; 95% CI = (.03, .17)), and was significantly correlated with study location (latitude; -.11, 95% CI = (-.16, -.06); longitude, .07, 95% CI = (.01, .13)).

In the bottom part of Figure 2 we present the posterior predictive medians and interquartile range of heritability, conditional on informant type (e.g., mother, father, teacher, observer, self) and sample decile values of age, and longitudinal studies (i.e., longsampl = 1), while controlling for all the other standardized moderators in the model by fixing them to zero. There, we see that the median and interquartile range of the heritability depend nonlinearly on age for each of the five types of informants, and that these dependencies are different across all types of informants. However, according to the interquartile range, the heritability distributions overlap among the five types of informant from early childhood through adolescence.

## Discussion

The three goals of the meta-analysis were to (a) infer the overall distribution of the heritability in the underlying population of studies, (b) identify significant moderators which served as significant predictors of changes in mean heritability, and (c) determine the extent to which key aspects of the heritability distribution (e.g., median, interquartile range) varied as a function of age and type of informant in longitudinal research. The meta-analysis revealed a bimodal overall heritability distribution in the underlying population of studies, indicating two clusters of heritability. Also, the analysis revealed four moderators that predicted significant changes in the mean heritability. Finally, the meta-analysis revealed differential patterns of median $h^2$ and variance (interquartile ranges) across informants and ages in longitudinal studies. These latter findings emerged from all studies in the present sample with a particular focus on longitudinal research, in which informants rated the same children and youth across one or more waves of data from early childhood through adolescence. Here we provide a general discussion of the results, followed by an analysis of their application to the study of informant discrepancies and behavioral genetics.

The first mode of distribution in the top portion of Figure 2 indicates one latent group of behavioral genetic studies with a median heritability of approximately .51. This heritability matches findings from previous meta-analyses of behavioral genetic research in antisocial behavior (i.e., Burt, 2009a; 2009b; Rhee & Waldman, 2002). The second mode of distribution indicates a second latent group of behavioral genetic studies with a median heritability of approximately .72. This heritability is considered to be very high (Loehlin, Neiderhiser, & Reiss, 2003). The second latent group of studies represented $n = 19$ of the 89 heritability estimates (21%); 8 of those 19 samples (42%) were represented by ratings from mixed informants.

In addition, we identified four significant moderators of (mean) heritability. Among them, heritability was on average significantly higher among studies having at least 60% white twin and non-twin siblings. In the current meta-analysis, we sought to examine the possible effect of ethnicity and SES, because in recent years behavioral genetic research has become more broad, recruiting individuals from diverse populations (examples from the present study include the National Longitudinal Study of Adolescent Health in the United States, the SIBS adoption study in Minnesota, the Taiwan twins study, and the USC twins). Studies from each of these data sets included 50% or fewer white twin and non-twin siblings in their samples. Although we did not obtain significant findings associated with SES in the present study, findings regarding ethnicity may be correlated with those for SES. That is, in previous behavioral genetic research, Turkheimer and colleagues found that in impoverished families, 60% of the variance in IQ was accounted for by the shared environment (and the contribution of genes close to zero), with the opposite finding for affluent families (Turkheimer, Haley, Waldron, D'Onofrio, and Gottesman, 2003). In the present study, we obtained a comparable finding regarding the heritability of antisocial behavior (significantly higher) among white families. Similarly, we found that heritability was significantly higher in the case of longitudinal compared to non-longitudinal research. Clearly, those who stay enrolled and continue to participate in longitudinal research are individuals who are more likely to be white, female, married, and with higher levels of education and better health—all proxies for higher SES (Radler & Ryff, 2010).

Finally, heritability was significantly correlated with study location. The slope coefficient estimates of latitude and longitude for the regression model (-.11, .07), respectively, showed that a one-unit increase in latitude corresponded to a decrease of -.11 in heritability on average, and that a one unit increase in longitude corresponded to decrease of .07 in heritability on average (after controlling for all other predictors and random effects in the model, by setting them to zero). Just as Turkheimer et al. (2003) found a significant relationship among heritability, SES, and IQ, we found clear associations between heritability of antisocial behavior and demographic factors (ethnicity, characteristics of participants in longitudinal research, and study location).

### Informant Discrepancies and the Prediction of Heritability

Fundamental questions about informant discrepancies have not changed since the seminal publication by Achenbach et al. (1987). Given that the selection of multiple informants is already a component of evidence-based practice in the assessment of child and adolescent mental health (Hunsley & Mash, 2007; Mash & Hunsley, 2005), and that research in behavioral genetics is heavily dependent on informants' views, how can research inform that selection? Which informant or combination of informants does one choose, and what is likely to be the effect on heritability? Already it appears from the present study that employment of multiple informants is associated with higher $h^2$. Should the combination of informants vary, depending upon the age at which youth are assessed, and the type of behavior (in the case of antisocial behavior, whether the behavior is predominately aggressive or rule-breaking)? Since the seminal Achenbach (1987) publication, researchers have advanced the study of informant discrepancies and their meaning (i.e., Bartels, Boomsma, Hudziak, van Beijsterveldt, & van den Oord, 2007; De Los Reyes, et al., 2009), even as research in the advanced stages of behavioral genetics proceeds, with results dependent upon the ratings of various informants.

We discuss patterns of median $h^2$ (and its variance, interquartile ranges) within and across the five informants employed in the present study. First, patterns of median $h^2$ looked very different across informants, with $h^2$ associated with mother, father, self, and observer ratings all declining from early childhood through adolescence, whereas $h^2$ associated with teacher ratings increasing during that same period. Similarly, we obtained different patterns of variance associated with the different informants, with the greatest variance in father and observer ratings over time. In each of the following sections, we focus on particular comparisons within and between raters in accordance with Kraemer et al.'s (2003) model of selecting informants, based on their differing perspectives and contexts, presented in Figure 3.

The goal of Kraemer et al.'s (2003) pragmatic framework is to assist investigators in selecting informants for their research by considering the total number of contexts and different types of perspectives about which informants can provide valid and reliable information (Kraemer et al., 2003). In the application of this framework, Kraemer et al. (2003) suggest that investigators divide the total context for behavior into two or more broad categories (i.e., home and nonhome) that are likely to influence informants' reports. The result is a mix-and-match approach to the selection of informants, contrasting those who hold the same perspective in different contexts (i.e., youth) with those who hold different perspectives in the same context (i.e., youth and teachers; youth and parents; Kraemer et al., 2003). We apply this framework to the discussion of our findings, focusing on median heritability results for informants and their corresponding variances (interquartile ranges) in Figure 2. Furthermore, we apply the framework to our discussion of implications for future research in the behavioral genetic study of heritability.

**Ratings by Informants in the Home: Mothers, Fathers, and Observers**

Mothers were clearly the most common informants in the present study. When more than one informant was recruited, fathers' ratings were frequently included alongside those of mothers, providing perspectives on behavior in the home setting. Although the overall trend in median heritability estimates from fathers (and to a lesser extent, mothers) declined over time, we observed distinct differences in interquartile ranges for the two informants. That is, ratings from mothers from early childhood through adolescence resulted in relatively small interquartile ranges; ratings from fathers resulted in relatively large interquartile ranges (second only to observers). In their meta-analysis of correspondence between mother and father ratings of externalizing behavior, Duhig, Renk, Epstein, & Phares (2000) found a high degree of correspondence between mothers and fathers in their ratings of externalizing behavior, confirming the same context/similar perspective provided by parents.

**Mother compared to father ratings.** Certainly differences between mothers and fathers could reflect a host of factors, including differences in the ways that mothers and fathers interact with children (Parke, 2000), the quality of the marital relationship (Hetherington & Clingempeel, 1992), degree of parent psychopathology (Connell & Goodman, 2002), and coherence of adolescent attachment (Ehrlich, Cassidy, & Dykas, 2011). Although ratings by parents can be the best predictor of child antisocial behavior in the home (De Los Reyes et al., 2009; Hartley et al., 2011; Ollendick, Jarrett, Wolff, & Scarpa, 2009), it may not be useful to rely solely on both mothers and fathers for ratings; nor may it be useful to add informants without considering those who can provide a different perspective on behavior in a different context. That is, informant discrepancies may reflect variation in behavior that is connected either to the setting (i.e., home and school) or to the informants' perspective on behavior (De Los Reyes et al., 2009).

**Parent compared to observer ratings.** In the present study, observers, both in the home and outside the home, were the least common informants, resulting in median heritability that declined slightly over time, and the largest interquartile ranges. This is likely because so few observer ratings were part of the present study, and each of the ratings was very different. Observer ratings in the home were provided in one study along with those of mothers, teachers, and young children themselves (Arsenault et al, 2003). In this study, observers completed a standardized assessment of aggression following a home visit of 2-3 hours in length. Interestingly, mothers and observers had the lowest levels of agreement among pairs of observers in the Arsenault et al., 2003 study, with correlations between pairs of informants as follows: mothers and observers, $r = .14$; mothers and teachers, $r = .28$; mothers and children, $r = .18$; teachers and observers, $r = .21$; and teachers and children, $r = .21$. The standardized assessment approach to ratings by observers can have predictive utility, predicting later engagement by youth in antisocial behavior, above and beyond family risk and parent ratings (Johnston & Murray, 2003).

**Ratings by Informants Outside the Home: Teachers and Observers**

After parents, teachers represented the next largest group of informants in the present study; their ratings were also likely to be combined with those of parents'. Teachers rate the behavior of youth within a relatively narrow and structured setting with clear expectations for behavior, and they are able to compare an individual student's behavior to that of his or her peers, representing a normative comparison group of the same age and gender (De Los Reyes et al., 2009). In these two ways (the narrowness of the context and the normative comparison), teacher ratings are very much unlike those provided by parents. In the present study, the overall trend in median heritability estimates from teachers increased over time, even as interquartile ranges became smaller.

Behavioral genetic researchers from the Twins Early Development Study (TEDS) and the Netherlands Twin Register have found that ratings by the same teachers of antisocial behavior among twins led to higher correlations between both MZ and DZ twin pairs than did ratings by different teachers (Polderman, De Sonneville, Verhulst, & Boomsma, 2006; Saudino et al., 2005). Correlations in both studies contributed to heritability estimates that differed, depending upon whether ratings by the same or different teachers were employed. Saudino et al. (2005) found significant differences in estimates of heritability based on same versus different teacher ratings for hyperactivity, prosocial behavior, and peer problems. Polderman et al. (2006) found that same teachers' ratings resulted in correlations nearly twice as large as different teachers' ratings, of both aggression and rule-breaking behavior, contributing to different heritability estimates.

**Teacher compared to parent ratings.** Loeber and colleagues have argued that teachers and parents are the optimal informants about hyperactivity and oppositional behavior, and children and parents should be employed as informants about CD (Loeber, Green, Lahey, & Stouthamer-Loeber, 1989; Loeber, Green, Lahey, & Stouthamer-Loeber, 1991). Although this approach takes into account perspectives on behavior, it does not fully account for settings in which behavior occurs.

Discrepancies between parents and teachers "emerge in part from the different behaviors they observe, attend to, and perceive as problematic within their perspective and setting" (Hartley et al. 2011, p. 57). Even though studies of informant discrepancies involving teachers have been relatively rare, in one longitudinal study, van der Ende and colleagues found that mean absolute differences between parent and teacher ratings of child and adolescent externalizing problems declined sharply from ages 4 to 17. This suggests that parents' and teachers' views may become more similar over time, with youth antisocial behavior perhaps occuring outside the home and school (van der Ende, Verhulst, & Tiemeier, 2012). In one study of a clinical sample, teachers did not deem their adolescent students to have clinically significant problems, even when those students had been hospitalized by their parents for psychiatric problems, confirming the context-specific nature of informants (i.e., outside school), and the limitations, in this case, of teachers' views (Talbott & Lloyd, 1997). Clearly, engagement in delinquent and rule-breaking behavior is more likely to occur outside the school and classroom during adolescence, and is most visible to youth and their peers.

**Teacher compared to observer ratings.** Outside the home in the present study, researchers employed observers using standardized criteria to evaluate observations in the laboratory during a card sort activity (Owen & Sines, 1970) and during family interactions (O'Connor, McGuire, Reiss, Hetherington, & Plomin, 1998). Clearly, we had a wide range of methods associated with observer ratings in the present study, which undoubtedly influenced the large interquartile ranges associated with observers. In the classroom setting, live observers are invaluable, but likely to identify low rates of externalizing behavior (given the setting), at a relatively high cost of conducting observations (Van Acker & Grant, 1996). On the other hand, De Los Reyes et al. (2009) successfully mapped the standardized assessment of observers in the laboratory to teachers' (but not parents') ratings of preschoolers' disruptive behavior. Although a structured laboratory setting does not match a year-long class placement, in which teachers develop relationships with students, the other-than-home context and the standardized perspective of an observer in De Los Reyes et al. (2009) suggests the utility of the cross-setting/cross-perspective framework.

**Ratings by Youth (Self) Informants in the Home and School**

After parents and teachers, youth ratings represented the next largest group of informants in the present study. Ratings by youth reflect their unique perspectives on their own behavior in various settings. That is, aggression and rule-breaking activities are increasingly likely to occur at times and places unobservable by parents or teachers as the child moves through adolescence. Across early childhood through adolescence, median $h^2$ associated with heritability for youth ratings declined, as interquartile ranges also became smaller.

**Youth (self) compared to parent ratings.** Parent and youth reports about externalizing behavior may be related to important aspects of the parent-child relationship and family functioning. For example, Ehrlich et al. (2011) found that among adolescents in a community sample, attachment coherence (the degree to which adolescents were securely attached to parents) was significantly and negatively correlated with adolescents' participation in externalizing behavior as reported by their peers. Comparably, discrepancies between parent and child ratings about externalizing behavior can be related to negative parenting practices, maternal stress, and the development of child behavior problems (Chi & Hinshaw, 2002; Ferdinand, van der Ende, & Verhulst, 2004; Pelton & Forehand, 2001). Yet, discrepancies between informants are not always the result of child impairment or observed problematic parenting; they may simply be the result of contexts for and informants' perspectives about behavior (De Los Reyes et al., 2009). In addition, when discrepancies arise between youth and parent ratings, clinicians and other practitioners may be more likely to assume that the parent is a correct and reliable informant about youth behavior, whereas youth are both incorrect and unreliable (De Los Reyes, Youngstrom, Swan, Youngstrom, Feeny, & Findling, 2011). How fair is this assumption? It may not be fair at all, especially when youth are engaging in delinquent or rule-breaking behavior, which is most likely to be out of the watchful eye of parents (Flannery, Williams, & Vazsonyi, 1999). However, in contrast to comparisons between youth and teacher ratings, in longitudinal assessments, mean absolute differences between youth and their parents about externalizing problems have declined gradually from ages 11 to 17 (van der Ende et al., 2012).

**Youth (self) compared to teacher ratings.** Unlike teachers, youth have access to observations of peers' and their own behavior in unstructured settings in the school. Rhee and Waldman (2002) noted that youth were most commonly, if not exclusively, employed as raters of antisocial behavior during adolescence. There are good reasons for this; at this age, youth are both aware of and cognitively capable of reporting about their own behavior. Antisocial behavior can spill over into the school setting in the form of disruptive and violent behavior. Youth do not necessarily agree with their teachers, however; in longitudinal assessments of ratings by youth and their teachers, mean absolute differences between youth and their teachers about externalizing problems increased sharply from ages 11 to 17 (van der Ende et al., 2012), reflecting the same perspective and context we discussed previously.

**Youth (self) compared to peers-as-observers ratings.** Peer nominations for aggression were collapsed with observer ratings in the present study (Brendgen, Dionne, Girard, Boivin, Vitaro, & Perusse, 2005), adding to the diversity of methods deemed "observations" in the present study. Peers are invaluable observers of youth behavior, particularly during adolescence, having direct access to youth participation in aggressive and delinquent behavior, substance use, and susceptibility to peer pressure, all in the absence of parental supervision (Flannery et al., 1999).

Clearly, research in the study of informant discrepancies has much to offer the study of behavioral genetics, given the emphasis in this research on ratings from parents, teachers, youth, observers, and other sources, such as school and criminal records. Continued progress in the study of informant discrepancies "will require a deeper understanding of how intwined behaviors and the surrounding social situation really are" (Hartley et al., 2011, p. 65). By the same token, behavioral genetics research has much to offer the study of informant discrepancies, by acting to disentangle informant bias from informants' unique perspectives about behavior (Bartels et al., 2007).

**Implications for Future Research in Behavioral Genetics**

The study of informant discrepancies in the context of behavioral genetics research is vital from both a scientific and a public health perspective. Kraemer (2011) describes a perfect methodological storm affecting the current study of genes interacting with environmental risk to predict mental health problems, with individual methodological challenges that are not new. Dependence upon informants' ratings as an outcome measure, without a clear understanding of how their perspectives differ (across ages and over time) clearly contributes to that storm. The goal of this pragmatic framework is to assist investigators in selecting informants for their research by considering the total number of contexts and different types of perspectives about which informants can provide valid and reliable information (Kraemer et al., 2003). Kraemer et al. (2003) suggest that investigators divide the total context for behavior into two or more broad categories (i.e., home and nonhome) that are likely to influence informants' reports. The result is a mix-and-match approach to the selection of informants, contrasting those who hold the same perspective in different contexts (i.e., youth) with those who hold different perspectives in the same context (i.e., youth and teachers; youth and parents; Kraemer et al., 2003). In the present meta-analysis, we found that relatively few studies employed the mix and match of informant perspectives and contexts advocated by Kraemer et al. (2003). There are indications from the present study that the use of multiple informants may result in higher heritability; of course, who those informants are and the perspectives and contexts they represent may also affect heritability.

Researchers in three studies in the present sample obtained the perspectives of three different informants in three different contexts, per the Kraemer et al. (2003) framework. These studies were conducted by Arsenault et al. (2003), employing the perspectives of mothers, teachers, youth, and observers; Eaves et al. (1997), employing the perspectives of mothers, fathers, teachers, and youth; and O'Connor et al. (1998), employing the perspectives of mothers, fathers, youth, and observers. Across these three studies and their five heritability estimates, the mean $h^2 = .54$, which is indeed in line with that reported in previous meta-analyses.

Results from previous research in the behavioral genetic study of antisocial behavior addressing the contributions of multiple informants have been mixed; each of these studies was included in our meta-analysis (Baker et al., 2007; Burt et al., 2005; Simonoff et al., 1995). Some of the studies found genetic influences to be higher in the case of multiple raters' shared views of disruptive, antisocial child behavior (Baker et al., 2007; Simonoff et al., 1995). For example, Baker et al. (2007) analyzed shared ratings of antisocial behavior provided by parents, teachers, and youth. Theirs was a study of 9 and 10 year old twins in southern California, whose data we have included in the present meta-analysis. Baker et al. (2007) found that the shared view of antisocial behavior was strongly influenced by genes, and that informants contributed differentially to that view, with proportions of variance contributed by parents being highest (.436), followed by teachers (.289), and then youth (.168), respectively. This highly heritable common factor of antisocial behavior was measured with data from multiple informants using the cross-setting, cross-perspective approach (Kraemer et al., 2003). Likewise, Simonoff et al. (1995), in the Virginia Twin Study, found genetic influences to be greater in the case of shared (father, mother, child) assessments of disruptive child behavior.

Burt and colleagues, using data from mother and child reports about 11 year old male twins, obtained the opposite finding. That is, when ratings from informants were combined to create a general externalizing factor (consisting of the common variance associated with ADHD, CD, and ODD), genetic factors did not account for a significant portion of the variance (Burt et al., 2005). Genetic factors were significant, however, when either mother or child informants' ratings were considered separately (Burt et al., 2005).

In the meantime, research at advanced levels of behavioral genetics proceeds. For example, a body of work has begun to accumulate in which researchers examine the gene that codes for the enzyme monoamine oxidase A (MAOA) interacting with the experience of maltreatment during childhood to affect antisocial and externalizing behavior (Kim-Cohen et al., 2006). Kim-Cohen et al. (2006) conducted a meta-analysis of five studies of this kind. In each of these studies, investigators examined the following G x E interaction: MAOA interacting with child maltreatment (obtained largely via retrospective accounts) to effect the emergence of antisocial behavior in adolescence and adulthood. In just two of the five studies analyzed by Kim-Cohen, researchers mixed and matched the perspectives of different informants in different contexts. In discussing their findings, these authors point out: "once an adverse experience touches off an otherwise 'silent' genetic vulnerability and triggers a cascade of biological events toward atypical development, what can be done to halt or reverse the process?" (Kim-Cohen et al., 2006, p. 911). The answer to that question can have tremendous public health implications, with possible pharmacological and environmental interventions the result of these and similar findings, all of which are ultimately dependent upon informants' views.

    In future work, researchers in behavioral genetics must do a better job of understanding the contributions of different informants to the study of child and adolescent psychopathology. Even as research in behavioral genetics proceeds, issues associated with informant discrepancies are just beginning to be sorted out (Achenbach, 2011). As a first step, behavioral genetics researchers can explore the effects of selecting a pragmatic model, such as that proposed by Kraemer et al. (2003). In the case of antisocial behavior, the model must be flexible enough so as to reflect the different developmental pathways that large numbers of youth are likely to follow from early childhood through adolescence (Loeber & Burke, 2011). Such will also be the case in the study of related externalizing disorders, such as ADHD and ODD, as well as internalizing disorders, which have measurement challenges unique to those disorders, as well as issues associated with informant discrepancies. Indeed, we argue for a cross-perspective, cross-setting model for selecting informants in behavioral genetic research that is flexible and sensitive to changes in antisocial behavior over time. Given the scientific and public health implications of behavioral genetic research in child and adolescent psychopathology, the time for developing a better understanding of informants' views and their contributions to research in behavioral genetics is now.

# References

***indicates study was included in the meta-analysis**

Informants and Heritability 2.6.13                                                                                                                                 56*Derks, E. M., Hudziak, J. J., Van Beijsterveldt, C. E. M., Dolan, C. V., & Boomsma, D. I. (2004). A study of genetic and environmental influences on maternal and paternal *CBCL* syndrome scores in a large sample of 3-year-old Dutch Twins. *Behavior Genetics*, *34*, 571-583.

*Dionne, G., Tremblay, R., Boivin, M., Laplante, D., & Perusse, D. (2003). Physical aggression and expressive vocabulary in 19-month-old twins. *Developmental Psychology*, *39*, 261-273.

Dodge, K. A., & McCourt, S. N. (2010). Translating models of antisocial behavioral development into efficacious intervention policy to prevent adolescent violence. *Developmental Psychobiology*, *52*, 277-285.

Duhig, A. M., Renk, K., Epstein, M. K., & Phares, V. (2000). Interparental agreement on internalizing, externalizing, and total behavior problems: A meta-analysis. *Clinical Psychology: Science and Practice*, *7*, 435-453.

DuMouchel, W. & Normand, S.-L. (2000). Computer-modeling and graphical strategies for meta-analysis. In D. Stangl & D. Berry (eds), *Meta-analysis in medicine and health policy*. Marcel Dekker, New York, pp. 127-178.

*Eaves, L.J., Silberg, J. L., Meyer, J. M., Maes, H. H., Simonoff, E., Pickles, A., Rutter, M., Neale, M. C., Reynolds, C. A., Erikson, M. T., Heath, A. C., Loeber, R., Truett, K. R., & Hewitt, J. K. (1997). Genetics and developmental psychopathology: 2. The main effects of genes and environment on behavioral problems in the Virginia twin study of adolescent behavioral development. *Journal of Child Psychology and Psychiatry,* 38, 965-980.

Informants and Heritability 2.6.13                                                                                    57

Informants and Heritability 2.6.13                                                                 58

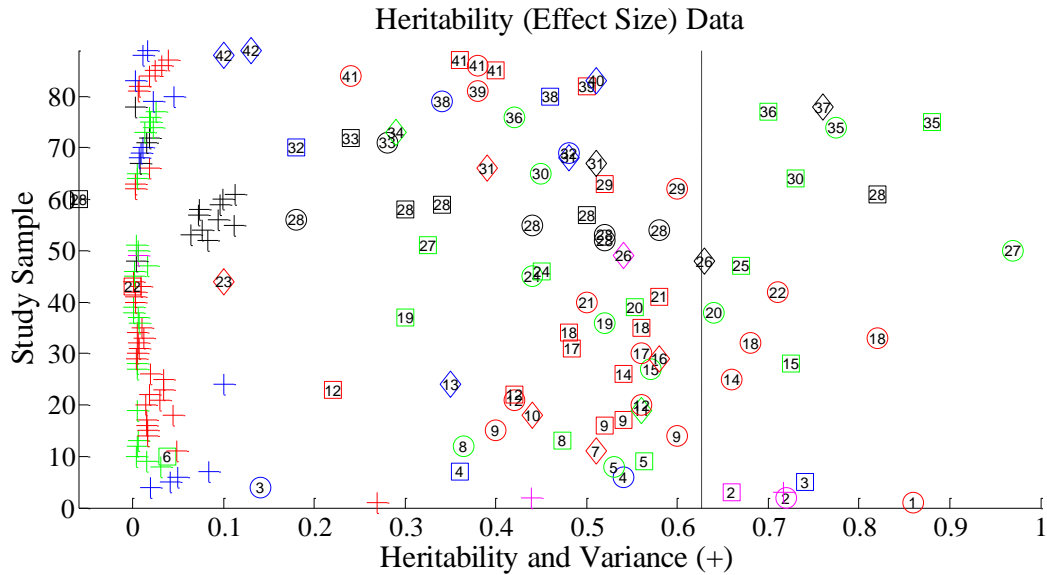

*Figure 1.* Heritability ($h^2$) and its variance (+) for $n$ = 89 samples provided by 42 studies. Of the samples, $n$ = 71 are from MZ and DZ twins (in 29 studies), for which we calcuated $h^2$ and $n$ = 18 are from studies of twin and non-twin siblings, and one adoption study (in 13 studies), wherein we retrieved $h^2$ from the original research. Shapes indicate gender, with circle=females; square=males. Diamond represents a heritability estimate from non-twin siblings, obtained via genetic model(s) which controlled for gender. Colors indicate the type of informant: red = mother only; black= teacher only; blue=self only; magenta=observer only; green = mixed informants (including fathers) with $h^2$_Ave=1. The dashed vertical line separates two latent clusters of studies with regard to heritability.

| Variable | Mean | S.D. | Type | Variable Description |
|---|---|---|---|---|
| $h^2$ | .48 | .20 | DV | Heritability $h^2$ estimate in a study. |
| Var_$h^2$ | .04 | .09 | | Var($h^2$) = $h^2$ sampling variance = 1/study weight. |
| SE | .15 | .13 | IV | Square root of Var($h^2$) |
| PY | 2001.1 | 9.2 | IV | Publication year. |
| $h^2$_Ave | .22 | .42 | IV | 1 if $h^2$ is Var($h^2$)-weighted mean of $h^2$ over informants. |
| female | .40 | .49 | IV | 1 if $h^2$ estimate from female siblings |
| Male | .42 | .50 | IV | 1 if $h^2$ estimate from male siblings |
| genmodel | .18 | .39 | IV | 1 if $h^2$ estimate controls for gender via a genetic model. |
| adoptee | .02 | .15 | IV | 1 if $h^2$ estimate from adoptees. |
| Mom | .50 | .44 | IV | 1 if mother ratings (proportion ratings if $h^2$_Ave=1). |
| Dad | .05 | .13 | IV | 1 if father ratings (proportion ratings if $h^2$_Ave=1). |
| teacher | .23 | .38 | IV | 1 if teacher ratings (proportion ratings if $h^2$_Ave=1). |
| Self | .18 | .36 | IV | 1 if self ratings (proportion ratings if $h^2$_Ave=1). |
| observer | .04 | .19 | IV | 1 if observer ratings (proportion ratings if $h^2$_Ave=1). |
| CD | .03 | .18 | IV | 1 if conduct disorder ratings (prop. ratings if $h^2$_Ave=1). |
| Agg | .37 | .48 | IV | 1 if aggression ratings (proportion ratings if $h^2$_Ave=1). |
| delinq | .09 | .28 | IV | 1 if delinquency ratings (proportion ratings if $h^2$_Ave=1). |
| Ext | .51 | .50 | IV | 1 if externalizing ratings (proportion ratings if $h^2$_Ave=1). |
| Achenb | .49 | .50 | IV | 1 if Achenbach questionnaire. |
| interview | .06 | .24 | IV | 1 if interview. |
| age | 120.9 | 49.8 | IV | Mean age of subjects in months. |
| white60 | .90 | .30 | IV | 1 if at least 60% whites in study. |
| zygquest | .73 | .45 | IV | 1 if zygosity obtained by questionnaire. |
| zygdna | .54 | .50 | IV | 1 if zygosity obtained by DNA samples. |
| sesmiss | .08 | .27 | IV | 1 if missing SES information. |
| seslow | .24 | .43 | IV | 1 if twins sample contains low SES subjects. |
| sesmidhi | .92 | .27 | IV | 1 if twins sample contains mid or high SES subjects. |
| repsample | .81 | .40 | IV | 1 if representative sample; 0 if convenience sample. |
| longsampl | .81 | .40 | IV | 1 if longitudinal sample; 0 if cross-sectional sample. |
| latitude | 46.8 | 8.7 | IV | Latitude of study. |
| longitude | −44.4 | 58.9 | IV | Longitude of study. |

*Table 1*. Descriptive statistics (*n* = 89) for the heritability estimate ($h^2$) dependent variable (DV), for the $h^2$ sampling variance (i.e., Var($h^2$)), and for the 29 moderators (IVs).

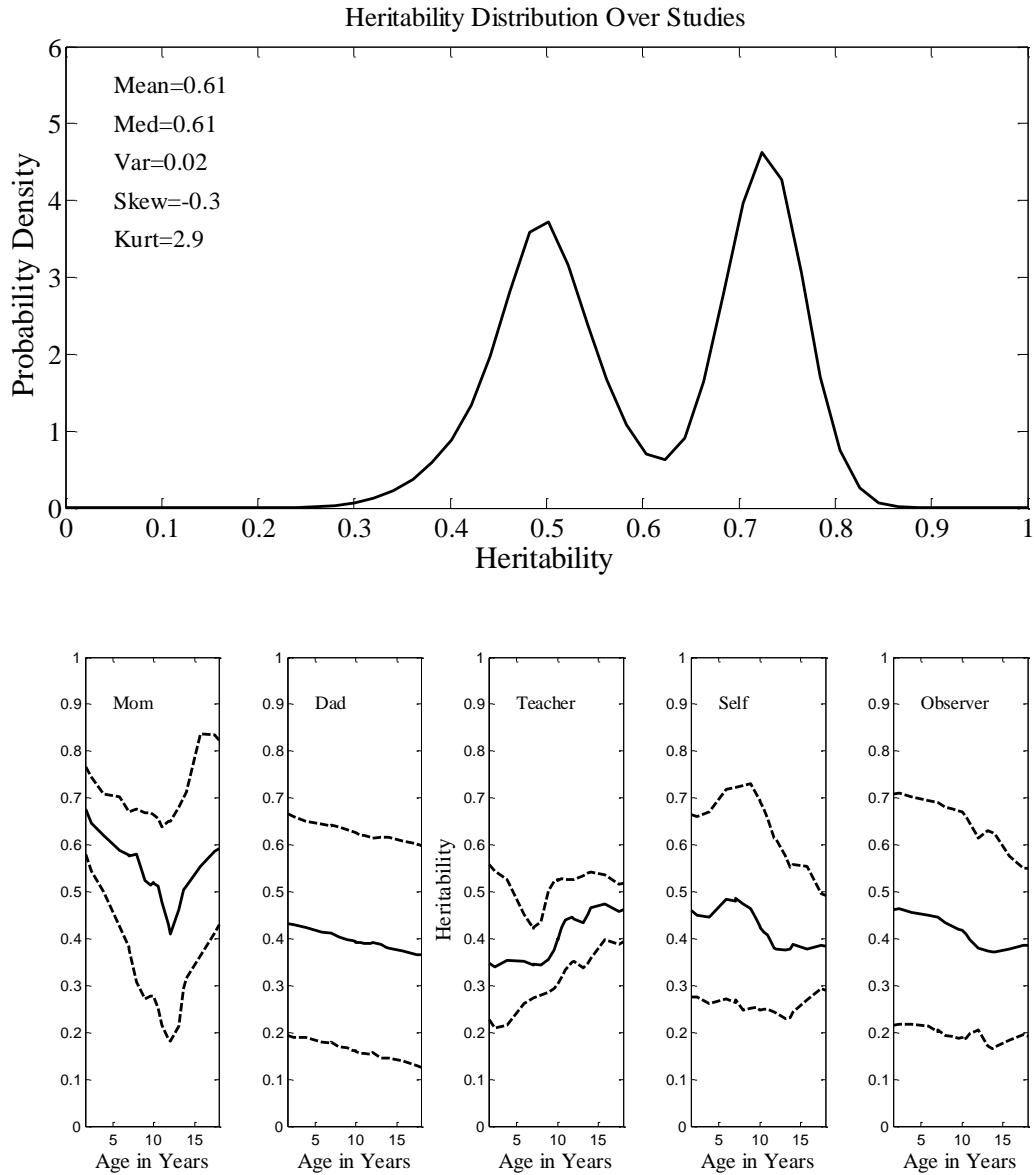

*Figure 2.* Top: From our Bayesian meta-analysis model, the posterior predictive density estimate of the heritability distribution for the underlying population of studies (Med=Median; Var=Variance; Skew=Skewness; Kurt=Kurtosis). Bottom plots: The median (solid line) and interquartile range (dashed lines) of the posterior predictive distribution of heritability ($h^2$), given informant type, age in years, for longitudinal studies, while controlling for all other predictors in the regression model.

| Moderator | Posterior Mean β | 95% CI | Moderator | Posterior Mean β | 95% CI |
|---|---|---|---|---|---|
| Intercept $\beta_0$ | .44 | (.39, .49) | delinq | .01 | (-.15, .17) |
| SE | .09 | (.02, .17) | ext | .04 | (-.31, .24) |
| PY | .01 | (-.05, .08) | Achenb | -.03 | (-.08, .02) |
| h2_Ave | .05 | (.01, .10) | interview | -.07 | (-.14, .00) |
| female | .01 | (-.25, .27) | age | -.02 | (-.06, .03) |
| male | -.02 | (-.28, .24) | white60 | .12 | (.06, .18) |
| genmodel | .01 | (-.20, .22) | zygquest | -.02 | (-.06, .02) |
| adoptee | .01 | (-.06, .07) | zygdna | -.04 | (.10, .02) |
| mom | .04 | (-.21, .30) | sesmiss | -.02 | (-.31, .28) |
| dad | -.02 | (-.11, .07) | seslow | .00 | (-.06, .05) |
| teacher | -.03 | (-.25, .19) | sesmidhi | .01 | (-.28, .31) |
| self | -.01 | (-.23, .20) | repsample | -.03 | (-.09, .02) |
| observer | -.01 | (-.12, .11) | longsampl | .09 | (.03, .17) |
| CD | .06 | (-.05, .17) | latitude | -.11 | (-.16, -.06) |
| agg | .01 | (-.25, .28) | longitude | .07 | (.01, .13) |

*Table 2*. For the linear regression coefficients, that is the intercept and the slope for each of the 29 moderators, the marginal posterior mean and 95% posterior credible (confidence) interval estimates.

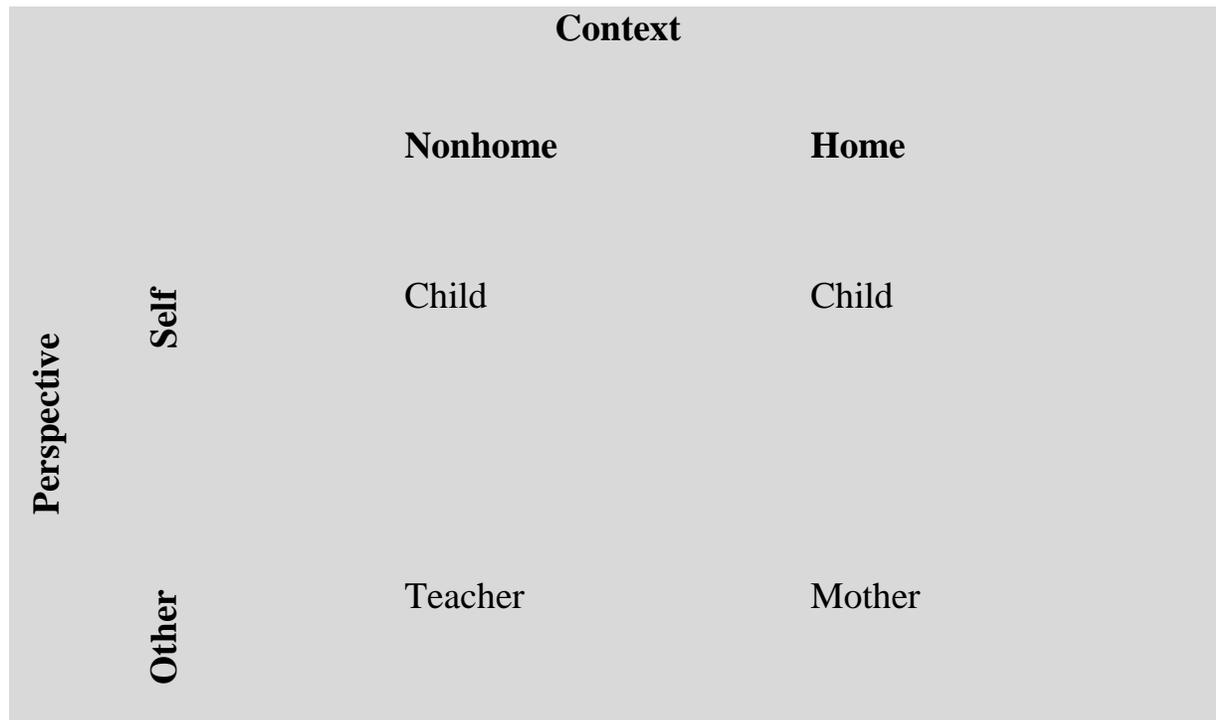

*Figure 3.* Multiple Informants Representing Perspectives and Contexts (Kraemer et al., 2003).